# Uncovering hidden macromolecular dynamics with Modulated Orientation Sensitive Terahertz Spectroscopy

Rohit Singh, Deepu Koshy George and Andrea Markelz
Department of Physics, University at Buffalo, Buffalo NY USA

*Abstract*—We introduce Modulated Orientation Sensitive Terahertz Spectroscopy (MOSTS) that can both increase contrast between adjacent modes and remove the broad band librational background to allow mode identification for large macro molecules such as proteins.

**Introduction**

For complex molecular systems, overlapping modes can interfere with determination of structure dependent resonances. In addition, librational excitations can overlap in energy with low frequency vibrational motions, giving rise to a strong smooth background. An example of such a system are proteins, where large scale correlated motions are predicted to lie in the THz range, however librational motions of the solvent and surface side chains interfere with the measurement of these functionally important large scale structural vibrations. We introduce Modulated Orientation Sensitive Terahertz Spectroscopy (MOSTS) that can both increase contrast between adjacent modes and remove the librational background. The technique is demonstrated using sucrose crystals and a model protein system. The sucrose results are in good agreement with a simple model. Further, the protein model system demonstrates that while static polarization dependent measurements do not reveal any sharp modes, the MOSTS measurements reveal clear narrow band resonances.

There have been many terahertz studies of vibrational modes in biomaterials in solution and crystalline forms [1-5]. The phonon modes in crystalline biomaterials such as sucrose, oxalic acid, benzoic acid and thymine [4, 6] etc. have been characterized. However for larger molecules the crystalline modes may be closely spaced and difficult to identify. Further, for large macromolecules such as proteins, vibrational modes that extend throughout the molecule will have frequencies in the terahertz range. In addition to a high density of modes, the energy range overlaps the librational excitations, which give a relaxational response with relaxation times ~ 1 ps. Thus the identification of specific structural vibrations for large macromolecules is problematic[7, 8].

The permittivity can be modeled as a sum of the global vibrational contribution and a local librational contribution.

$$\varepsilon(\omega) = \varepsilon_\infty + \sum_{i=7}^{3N-6} \frac{f_i}{\omega_i^2 - \omega^2 - i\omega\gamma(\omega_i)} + \varepsilon_r \int_0^\infty \frac{h(\tau)d\tau}{1+i\omega\tau} \quad (1)$$

Where $\varepsilon_\infty$, $f(\omega')$, $g(\omega')$, $\gamma$, $\tau$, and $h(\tau)$ are the permittivity at high frequency, the oscillator strength for resonant frequency $\omega'$, the density of states for frequency $\omega'$, the damping of mode $\omega'$, relaxation time and the distribution of relaxation times.

In order to improve contrast between modes and suppress the relaxational contribution to the signal, we can utilize the fact that the coupling to a protein collective mode is sensitive to the protein's orientation relative to light polarization. Figure 1 shows a cytochrome-c. The arrows indicate the atomic displacements associated with the vibrational mode at 9.72 cm$^{-1}$. The net dipole change for the vibration is along the direction shown by the red arrow. Excitation of the mode is maximum for 9.72 cm$^{-1}$ light with polarization parallel to this vector, where if the light has the perpendicular polarization the mode will not be excited and no absorption will be detected.

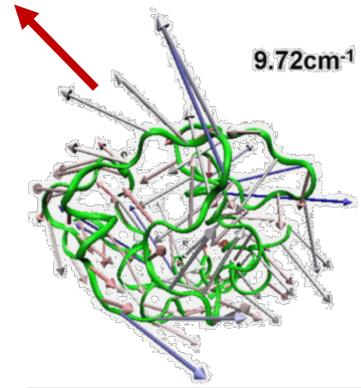

Figure 1. Displacement vector diagram of the eigenvector associated with an eigenmode of cytochrome-c at 9.72 cm$^{-1}$. The red vector shows the net direction of the change in dipole moment with the mode.

The absorption is dependent on the dot product of the light polarization and the dipole derivative for a particular mode.:

$$Abs_{aligned} = \sum_{i=7}^{3N} \frac{\gamma^2/\omega_i}{(\omega-\omega_i)^2+\gamma^2}\left[\left(\frac{\partial \vec{p}}{\partial q_i}\right)\cdot\hat{\lambda}\right]^2 \quad (2)$$

Thus polarization dependent measurements on aligned samples will enable separation of closely spaced modes through their different orientation dependence. More importantly the relaxational contribution will not have this orientation dependence. The hydration water is isotropically oriented over the protein surface and interior. Similarly the side chains have dipoles with directions homogeneously distributed over the protein. Using polarization sensitive measurements then on aligned samples should both allow removal of the relaxational background and mode separation. The most straight forward method to achieve this is by making multiple measurements for the aligned sample rotated relative to the light polarization and calculating the change in the absorption for different rotations. The disadvantage of performing these static measurements is that they are dominated by a large relaxational background, which is only removed after subtraction with different rotations. Thus one is measuring a small signal on a large background. A better approach is to modulate the polarization relative to the crystal orientation and measure only the signal dependent on the modulation. Previously we demonstrated high sensitivity measurements of Faraday and Kerr angles by modulating the incident polarization for solid state systems such as topological insulators and two-dimensional electron gases realized in GaAs/AlGaAs heterostructures. The polarization modulation technique reduces the THz signal level, and is not ideal for linear dichroism measurements. Here we present an alternative method, where the orientation of the sample is modulated: Modulated Orientation Sensitive Terahertz Spectroscopy (MOSTS).

A standard THzTDS a photoconductive antenna generation and electro optic detection are used to generate and detect THz [9]. We use a Ti:Sapphire laser output with 100 fs pulse width and centered at 800nm (NIR). The 400 mW NIR beam is split into two beam paths: a THz generation path (350 mW) and NIR detection path (50 mW). THz light is generated by a photoconductive GaAs antenna with 80µm electrode separation which is DC biased (60V). The THz passes through the sample and is detected using electro optic detection using 2 mm thick (011) ZnTe crystal. Unlike standard THzTDS, in MOSTS the orientation of the sample is modulated by spinning it. The spin frequency of the sample is monitored and is used as the reference for lockin detection of the EO signal. By measuring only the signal at the spin frequency, we measure only the change in THz transmission with sample orientation.

**Modeling of MOSTS signal**. We can use Jones matrices to determine the field measured using the MOSTS technique. Let $E_i(\nu)$ be the field amplitude of the linearly polarized generated THz at frequency $\nu$. To determine the transmission of the birefringent sample, we rotate the THz into the sample's frame with ordinary and extraordinary axes. We then propagate the beam through the birefringent crystal, assuming no off diagonal components (that is no circular dichroism). We then rotate back to the lab frame. Finally, we use a final polarizer to ensure that we only detect non rotated light. The transmitted field $E_t$ can be determined by the Jones matrices corresponding to this sequence of optical operations are:

$$E_t = \begin{pmatrix} 1 & 0 \\ 0 & 0 \end{pmatrix} \begin{pmatrix} \cos\theta & \sin\theta \\ -\sin\theta & \cos\theta \end{pmatrix} \begin{pmatrix} e^{i2\pi\nu d\sqrt{\varepsilon_o(\nu)}} & 0 \\ 0 & e^{i2\pi\nu d\sqrt{\varepsilon_e(\nu)}} \end{pmatrix}$$
$$\times \begin{pmatrix} \cos\theta & -\sin\theta \\ \sin\theta & \cos\theta \end{pmatrix} \begin{pmatrix} E_i(\nu) \\ 0 \end{pmatrix} \quad (3)$$

$$= \frac{E_i(\nu)}{2} \begin{pmatrix} \left(e^{i2\pi\nu d\sqrt{\varepsilon_o(\nu)}} - e^{i2\pi\nu d\sqrt{\varepsilon_e(\nu)}}\right)\cos(2\omega t) + \left(e^{i2\pi\nu d\sqrt{\varepsilon_o(\nu)}} + e^{i2\pi\nu d\sqrt{\varepsilon_e(\nu)}}\right) \\ 0 \end{pmatrix}$$

where $\theta, d, \varepsilon_o, \varepsilon_e$ are the angle the polarization makes with the ordinary crystal axis, the sample thickness, the ordinary permittivity and the extraordinary permittivity respectively. For a sample rotated at angular velocity $\omega$, $\theta = \omega t$. If we use $2\omega$ as our lock-in reference, our net signal is given by:

$$E_{MOSTS}(\nu) = \frac{E_i(\nu)}{2}\left(e^{i2\pi\nu d\sqrt{\varepsilon_o(\nu)}} - e^{i2\pi\nu d\sqrt{\varepsilon_e(\nu)}}\right) \quad (4)$$

The broad band THz pulse has multiple frequency components, which can be determined by Fourier transform of the incident pulse. One can then model the net MOSTS signal in the time domain using an inverse Fourier transform:

$$E_{MOSTS}(t) = \int_0^\infty E_{MOSTS}(\nu) e^{i\nu t} d\nu \quad (5)$$

**Calibration with molecular crystals**. We demonstrate MOSTS using a well characterized anisotropic system, sucrose. Large sucrose crystals were made using the seed growth method. Sucrose powder were purchased from Sigma Aldrich. Sucrose was dissolved in pure DI water to saturation at 23°C. Small sucrose seeds were first grown by drying saturated solution. Seeds were then hung in a beaker filled with clean saturated solution and were left for several days at room temperature. Sucrose crystals as large as 15 mm x 12 mm x 7 mm were obtained and habits were sufficiently clear that one could readily identify the crystal axes. The grown crystals were polished by hand along a specific crystal face (a, b or c) to have a large area and be sufficiently thin so that the absorbance is within the dynamic range of the THz system.

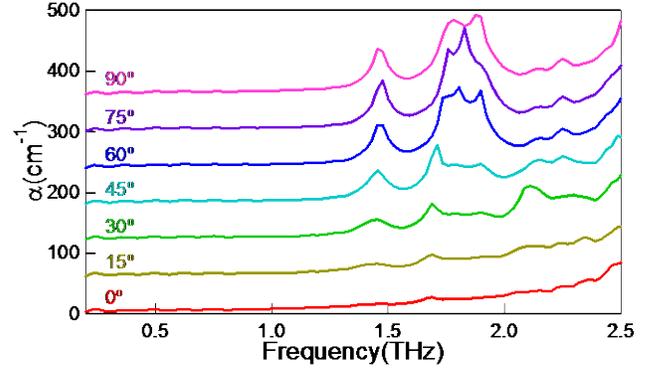

Figure 2. Orientation dependence of absorbance spectra and for a 550µm thick a-cut sucrose crystal. 0° corresponds to E||b crystal axis.

Standard THz TDS measurements of the anisotropic response of a 650 µm thick a-face sucrose crystal are shown in Fig. 2. The strong THz absorption is highly dependent on the crystal orientation relative to the incident THz polarization, with particularly strong features at 1.47 THz, 1.9 THz and 2.1 THz. THz index measurements find that sucrose is slightly birefringent with $\Delta n = 0.05$. We use these measurements in the modeling in Eq. 3 and Eq. 4.

We use a 1 mm thick sucrose crystal in the MOSTS system. The results are shown in Figure 3. We show both the calculated and measured MOSTS waveforms. We note, when an isotropic sample such as silicon is measured there is no detected waveform. There is good agreement between the calculated and measured MOSTS waveforms. We also show the calculated and measured MOSTS power spectrum. The frequency dependent features are in good agreement. The absorption peaks correspond to inflection points in the MOSTS power spectrum. The sucrose data demonstrate that MOSTS is sensitive to the anisotropic response, but also demonstrate that MOSTS is not an ideal method to characterize anisotropic crystals that do not have a large relaxational background. The static THz TDS measurements are better for this type of characterization. But our goal is to see anisotropic response for crystals with large relaxational background. We will now demonstrate that MOSTS does indeed excel at this, enabling the determination of anistropic response that static measurements cannot reveal.

In the ideal case, we would demonstrate this directly with hydrated protein crystals. Unfortunately this requires a protein crystal size in excess of our diffraction limited spot size of 2 mm FWHM. Such large crystals are difficult to obtain at this time. In addition, the hydrated protein crystal needs to be mounted to the spinning sample mount in a controlled hydration environment. As these technical difficulties are being considered, we first demonstrate the MOSTS technique can measure anisotropic response above a large relaxational background by using a model system. We require a system that has both a large glass-like relaxational background and a strong anisotropic response. We create such a sample by attaching a sucrose crystal to a thick polycarbonate (PC) substrate. PC has a glass-like response as seen in Fig. 4. The PC substrate give the same large relaxational background as we get from the solvent and surface side chains for proteins, whereas the sucrose will give the large anisotropic vibrational response we expect from the global structural motions of the protein and depicted in Fig. 1. The model sample then consisted of a 2.99 mm thick PC substrate and a 0.56 mm thick c-cut sucrose crystal.

**Protein Model System Results.** Figure 5 shows the static measurements on the model sample along with the MOSTS

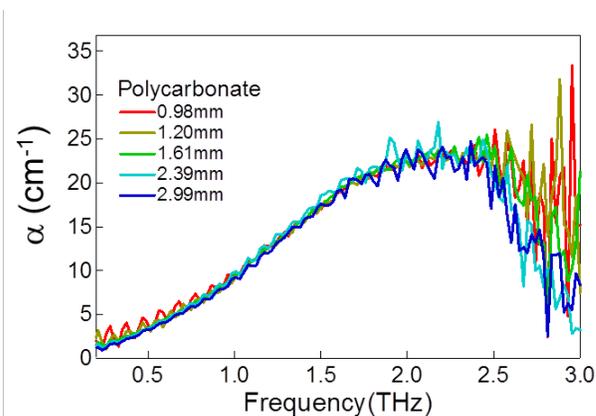

Figure 3. Absorption coefficient for polycarbonate measured for several sheets of varying thickness.

measurements. Plot (a) is the absorption spectra as a function of frequency for orientation 0° and 90°. There are no sharp resonances in this plot. It is just a broad glassy background. We cannot see the expected anistropic resonance at 1.47 THz. When we subtract the absorbances at these two orientations we see only noise as shown by plot 5(b). The sensitivity of the measurement, and the requirement of seeing a small change on a large background demonstrates that static measurements are not sufficient to see the anistropic absorbance above a strong relaxational background.

Turning to the MOSTS data, Fig. 5(c) shows the MOSTS waveform for the model sample. In spite of a large isotropic background, a MOSTS waveform is readily measured. Further, the MOSTS power spectrum in 5 (d) indicates there are anisotropic resonances in the 1.5 THz and 2.0 THz regions. The difference spectra from the static measurements do not indicate any strong spectral features at these frequencies. We note the similarity of the power spectrum for the large relaxational background sample to the pure single crystal data shown in Fig. 4 (d). The MOSTS method is highly successful at directly detecting anistropic response above a strong broad isotropic background.

## Conclusion

MOSTS is an effective technique to suppress glassy background for oriented molecular systems. IT gives spectroscopic information for molecular crystal and suppresses large glassy background for protein crystal model system

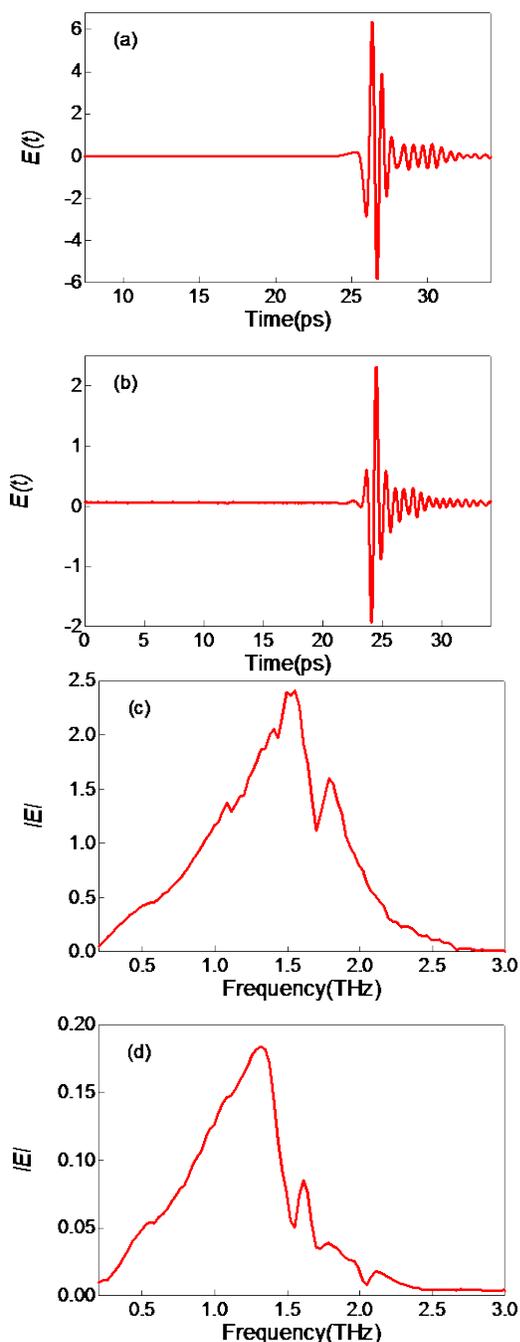

Figure 3 a) Calculated (b) Measured MOSTS waveforms (c) Calculated and (d) Measured power spectra of an A cut sucrose

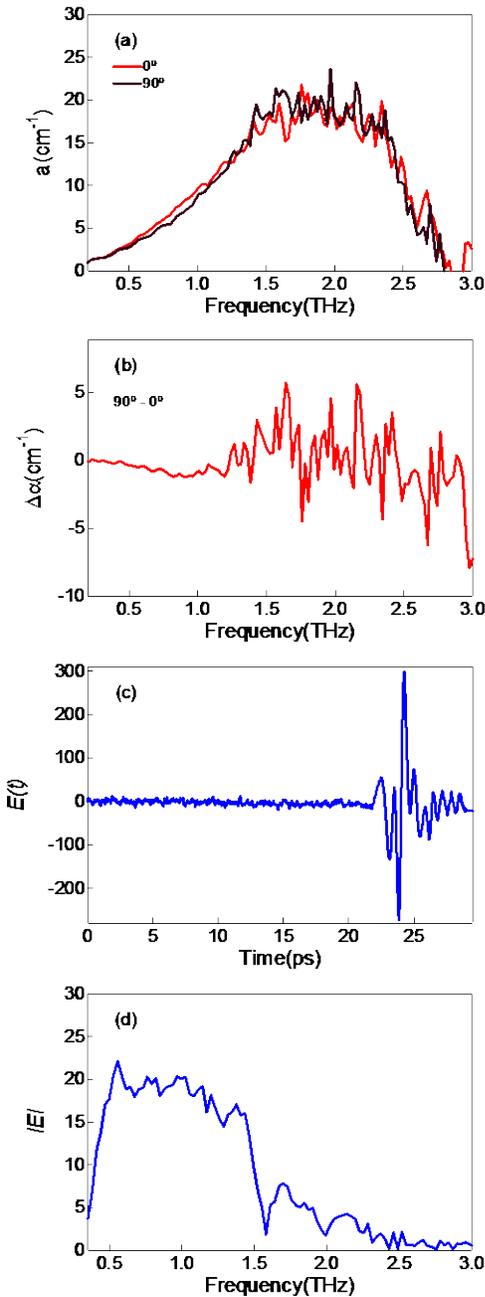

Figure 5. Measurements of a protein crystal model sample demonstrating that while static measurements are unable to detect anisotropic response above a large relaxational background, the MOSTS method is able to do so. (a) absorption spectra (b) difference between absorption spectra for the two orientations (c) MOSTS waveform (d) MOSTS power spectra of the model sample.